# Augmented Reality Technology in Teaching about Physics: A systematic review of opportunities and challenges

*Andrej Vidak[1], Iva Movre Šapić[1], Vanes Mešić[2], Vjeran Gomzi[3]

[1]Faculty of Chemical Engineering and Technology, University of Zagreb, Marulićev trg 19, 10000 Zagreb, Croatia
[2]Faculty of Science, University of Sarajevo, Zmaja od Bosne 35, 71000 Sarajevo, Bosnia and Herzegovina
[3]Faculty of Electrical Engineering and Computing, University of Zagreb, Unska 3, 10 000 Zagreb, Croatia

*Corresponding author email: avidak@fkit.unizg.hr

The use of augmented reality (AR) allows for the integration of digital information onto our perception of the physical world. In this article, we present a comprehensive review of previously published literature on the implementation of augmented reality in physics education, at the school and the university level. Our review includes an analysis of 96 papers from the Scopus and Eric databases, all of which were published between January 1st, 2012 and January 1st, 2023. We evaluated how AR has been used for facilitating learning about physics. Potential AR-based learning activities for different physics topics have been summarized and opportunities, as well as challenges associated with AR-based learning of physics have been reported. It has been shown that AR technologies may facilitate physics learning by: providing complementary visualizations, optimizing cognitive load, allowing for haptic learning, reducing task completion time and promoting collaborative inquiry. The potential disadvantages of using AR in physics teaching are mainly related to the shortcomings of software and hardware technologies (e.g., camera freeze, visualization delay) and extraneous cognitive load (e.g., paying more attention to secondary details than to constructing target knowledge).

KEYWORDS: augmented reality, mixed reality, assisted reality, physics education, systematic review.

## 1. Introduction

The history of physics shows that deeper truths about the physical world are often hidden under the surface of our immediate experience. Therefore, for successful learning of physics it is beneficial to extend the learner's immediate experience of physical phenomena. To that end the learner may be immersed into different kinds of environments and realities. Realities that allow for experiences that go beyond the immediate real-world experience can be provided through the use of modern technologies. A powerful classification of such realities has been recently provided by Rauschnabel *et al.* in their XReality framework [1]. According to this framework, if the real-world environment is at least visually part of the learning experience, then this kind of computer-enhanced reality is defined as augmented reality (AR) and otherwise it is defined as virtual reality (VR). In other words, augmented reality simply "represents a combination of real and virtual content that is displayed in real time" [1, p.7], whereas in virtual reality at least the visual aspects of the real-world environment are completely replaced by computer-generated information. In the XReality framework, assisted reality and mixed reality are seen as just two subtypes of augmented reality: the main difference between these two is that assisted reality is characterized by text-based instructions overlaid on the physical environment and mixed reality is characterized by seamless integration of (more complex) virtual and real objects. Within the XReality framework the term extended reality (XR), which has been earlier used by some authors as an umbrella term for AR, MR and VR, is suggested to be completely avoided because VR does not extend the real-world environment, but it completely replaces its visual aspects.

The first known head-mounted augmented reality (AR) system was invented in 1968 by Ivan Sutherland, an associate professor of electrical engineering at Harvard University [2]. In the 1970s and 1980s, renowned American laboratories and universities worked to advance AR technology, while in the 1990s AR was used to train pilots [3,4]. By 2022 augmented reality technology has been used in many areas of science, technology and industry such as Medical Training, Retail, Repair & Maintenance, Design & Modeling, Business Logistics, Tourism Industry, Education [5].

In the early days, AR systems often involved expensive electronic hardware (e.g., head-mounted displays), but nowadays AR is becoming more popular as augmented reality applications are supported by personal computers and mobile devices [6,7]. Some auxiliary hardware technologies commonly used in AR applications are [8-10]: smartglasses (i.e., wearable computer glasses with a camera and digital screen), motion sensors (i.e., devices that detect changes in movement), and projectors (i.e., devices that project digital content onto the real world, e.g. walls).

AR applications are developed using software development kits (SDK) such as Unity and Vuforia [11,12]. There are several different types of AR applications available, including marker-based, location-based, motion-based, and markerless AR [13]. Marker-based augmented reality uses a camera and designated printed markers to activate the AR content, while location-based AR uses information about the user's geographic coordinates to provide location-specific AR content. Motion-based AR is activated by changes in movement, providing a responsive and interactive experience. Markerless AR, on the other hand, is activated without the need for external triggers such as printed markers, utilizing techniques like image recognition or facial recognition.

## 1.1. Augmented reality technology in education

### 1.1.1. Relevant learning theories

Theories that are considered relevant for reasoning about learning with visualizations are also relevant for AR-based learning. Euler *et al.* [14] consider the following theories as relevant for learning with visualizations: cognitive load theory, theory of distributed cognition, cognitive theory of multimedia learning, Ainsworth's theory of multiple representations, grounded and embodied cognition, and social semiotics.

According to cognitive load theory [15], the working memory is very limited in capacity and duration. It may become overloaded which impedes learning. Concretely, learning environments may induce three mutually additive types of cognitive load: germane load (i.e., load resulting from learning processes that are relevant for constructing target knowledge), extraneous load (i.e., load related to information that is not relevant to learning) and intrinsic load (i.e., load related to inherent complexity of the learning task). For learning to be effective germane load should be fostered, extraneous load should be minimized and intrinsic load should be adjusted (e.g., by segmenting information) [16]. How this may be achieved effectively in learning with multimedia is nicely described by the Mayer's cognitive theory of multimedia learning which includes twelve principles. Particularly relevant for AR-based learning are the multimedia principle, principle of spatial contiguity, temporal contiguity, and the segmenting principle [17].

The multimedia principle states that people learn better from words and pictures than from words alone. AR tools can be designed to combine words, images, 3D objects, animations and movies, e.g., by integrating animations and videos into textbooks or by combining informative texts with digital images.

The principle of spatial contiguity states that people learn better when words and corresponding images are presented close to each other rather than distant from each other. AR tools provide the opportunity to place words and images near animated or physical objects. For example, we can overlay computer-generated textual descriptions on top of real-world car engine parts. Similarly, during a science experiment, we can display live measurement data on the AR glasses.

The principle of temporal contiguity states that people learn better when words and corresponding images are presented simultaneously rather than sequentially, while the segmenting principle states that people learn better when we divide a multimedia lesson into self-paced segments. AR applications could easily be reconciled with these two principles because they may simultaneously show words and images, and one can design them to consist of multiple mutually related, conceptual chunks.

Other theories that are relevant for AR-based learning are the theories of embodied and grounded cognition, as well as the theory of distributed cognition. According to the theories of embodied and grounded cognition "cognitive activity is grounded in sensory–motor processes and situated in specific contexts and situations" [18, p.1]. For example, AR technologies allow changing distance between two simulated physical bodies by physically moving the corresponding AR markers or affecting simulated meteor motion by moving through the classroom. The theory of distributed cognition asserts that cognition is distributed across "internal human minds, external cognitive artifacts, and groups of people" [19, p.333]. Due to the fact that processing of information occurs in a coupled system of internal and external representations, external representations such as AR-visualizations facilitate learning by preventing cognitive overload.

AR may support learning by displaying multiple representations on top of the corresponding real-world scenery. A systematic overview of how external representations may facilitate learning has been provided by Ainsworth. Ainsworth [20] points out that multiple representations can support learning in three ways:

- By allowing for complementary information or processes (e.g., light intensity distribution graphs and image of diffraction pattern: the image gives us insight into the shape of the diffraction fringes and the graph provides more detailed information about the distribution of light intensity)
- By limiting the room for interpretation (e.g., video of the motion and corresponding real-time graph: the video of the motion potentially prevents the interpretation of graph as the image of motion)
- By providing the opportunity for constructing deeper understanding (e.g., promoting relational understanding by associating x-t with the $v_x$-t graph)

If not introduced carefully, learning with multiple representations may result in cognitive overload.

Finally, taking into account that AR technologies often spark vivid classroom discussions, AR-researchers may also find it interesting to use the theoretical perspective of social semiotics [21], i.e., to investigate how meaning-making resources (e.g., talk, gestures, body position, disciplinary representations) are combined for meaning-making within the context of AR-learning situations.

### 1.1.2. Previous literature reviews

Some of the most prominent reviews on AR in education that have been published in last ten years are presented in table 1. Although many systematic reviews about AR in education were published, only a few of them were focused on AR in STEM education. The number of included articles ranged from 12 in the science education study by Cheng and Tsai [22] to 68 in the general education study by Akçayir *et al.* [23]. The most recent systematic review on AR in education was published in 2022 and focused on AR applications in STEM education; it included 45 articles [24].

Table 1. List of past AR education review studies.

| Authors | **N[a]** | **Research focus** |
|---|---|---|
| Cheng and Tsai (2012) [22] | 12<br>SCE | Technology features, Learning process and outcomes, Learning experience and interaction, Learner characteristics |
| Bacca *et al.* (2014) [25] | 32<br>GE | Uses, Purposes, Advantages, Limitations, Effectiveness and affordances, Adaptive or personalized process, Evaluation methods, Special needs of students |
| Akçayir *et al.* (2017) [23] | 68<br>GE | Papers per year, Types of participants, AR technologies, Advantages, Challenges |
| Chen *et al.* (2017) [26] | 55<br>GE | Papers per year, Number of publications per journal, Papers per country, Level of education, Research field, Research methods and design, Effectiveness |
| Ibáñez and Delgado-Kloos (2018) [7] | 28<br>SE | Educational level, Educational context, Sample size, Technology features, Instructional strategies and techniques, Research methods, Learning outcomes, Reported problems |
| Alzahrani (2020) [27] | 28<br>GE | Research design, Participants' education level and sample size, Benefits, Challenges |
| Sırakaya and Sırakaya (2020) [28] | 42<br>SE | Papers per year, Participants' education level and sample size, Data collection tools, Advantages, Challenges |
| Mystakidis *et al.* (2022) [24] | 45<br>SE | STEM fields, Software and hardware equipment, Design elements and visual features, Instructional methods |

[a] N = number of reviewed articles, GE = general education, SE = STEM education, SCE = science education.

Cheng and Tsai [22] identified major approaches to utilizing AR technology in science learning. They pointed out that marker-based AR supports students' spatial ability, practical skills and conceptual understanding, while location-based AR mainly supports inquiry based science activities.

Bacca *et al*. [25] found that majority of earlier AR research was situated within the context of science, humanities and arts topics. They emphasized that AR promotes learning gains, motivation, interaction, and collaboration. Longitudinal AR studies were recommended for purposes of investigating evolution of knowledge and skills over time.

Akçayir *et al*. [23] identified several advantages and challenges related to augmented reality for education. Most of the identified AR advantages were related to enhanced learning (i.e., engagement, interest, knowledge gain, satisfaction), while the challenges imposed by AR were related to usability issues and technical problems.

Chen *et al*. [26] reported on AR research in educational settings considering uses, advantages, characteristics, and effectiveness of AR in a variety of fields (science, engineering, health, social sciences). They concluded that the number of studies has increased significantly since 2013 and that most empirical studies have been conducted on science topics.

Ibáñez and Delgado-Kloos [7] particularly focused on AR instructional strategies and techniques. They found that most AR applications include simulations or exploratory activities and provide mainly visual learning experiences. They also found that AR could often lead to cognitive overload.

Alzahrani [27] conducted a systematic review of augmented reality in general e-learning contexts. He concluded that AR facilitates tactile, collaborative, creative, and distance learning. In addition, AR was found to enhance students' motivation, engagement, interactivity, concentration and knowledge retention while challenges were associated with cognitive overload and technical issues.

Sirakaya and Sirakaya [28] presented a systematic review of AR applications in STEM education. They reported that most of the research was conducted with K-12 students in the school environment. Furthermore, they identified better educational outcomes as one of the key advantages of AR, while teacher resistance and technical problems were recognized as main disadvantages.

In the study by Mystakidis *et al*. [24] three common augmentation methods were identified, including the enhancement of laboratory equipment, physical objects, and course materials. Additionally, the literature review noted various instructional strategies, software, and hardware used in augmented reality and found that students who used AR in STEM had improved outcomes compared to those who received traditional instruction.

It can be concluded that some earlier AR review articles included physics education studies. However, in our opinion the earlier published reviews missed providing an overview of AR-based *learning activities* within the context of individual physics topics. The systematic review presented in this article can make an important contribution to the current literature because it describes pedagogical, as well as technical details related to the use of AR in the learning and teaching about *individual physics topics*. We hope that the provided overview of learning activities may spark the physics teachers' interest in AR technologies and make it easier for them to effectively use AR applications in their own practice. In addition, this review presents multiple methodological aspects of earlier AR physics education research, which may provide the physics education researcher with useful information on research gaps and ways to improve AR research methodology.

### 1.2. Study purpose and research questions

The goal of this review is to provide a comprehensive overview of opportunities and challenges in AR-based teaching and learning about physics topics.

We aimed to answer the following research questions:

**RQ1:** How are augmented reality technologies used for facilitating learning about individual physics topics?
*Significance: Answering this research question could help us vividly demonstrate the inherent features of AR technology that can facilitate learning in the context of different physics topics. This summarizing of AR-based learning activities within the context of individual topics could inspire physics teachers to use AR in their own practice.*

**RQ2:** What are the specific advantages and disadvantages of using AR in teaching about physics topics?
*Significance: Answering this research question could help us generate research-based claims about advantages and disadvantages of using AR when teaching about various physics topics. Thus, we could draw conclusions about the consistency of earlier research and we could gain insight into the effects of AR in different domains of learning.*

## 2. Methods

In order to provide a valid answer to our research questions, we firstly had to identify a pool of high-quality articles related to AR-based learning and teaching about physics.

Our article selection procedures followed the guidelines provided by PRISMA (Preferred Reporting Items for Systematic Reviews and Meta-Analyses) [29]. The workflow for our article selection process is presented in figure 1.

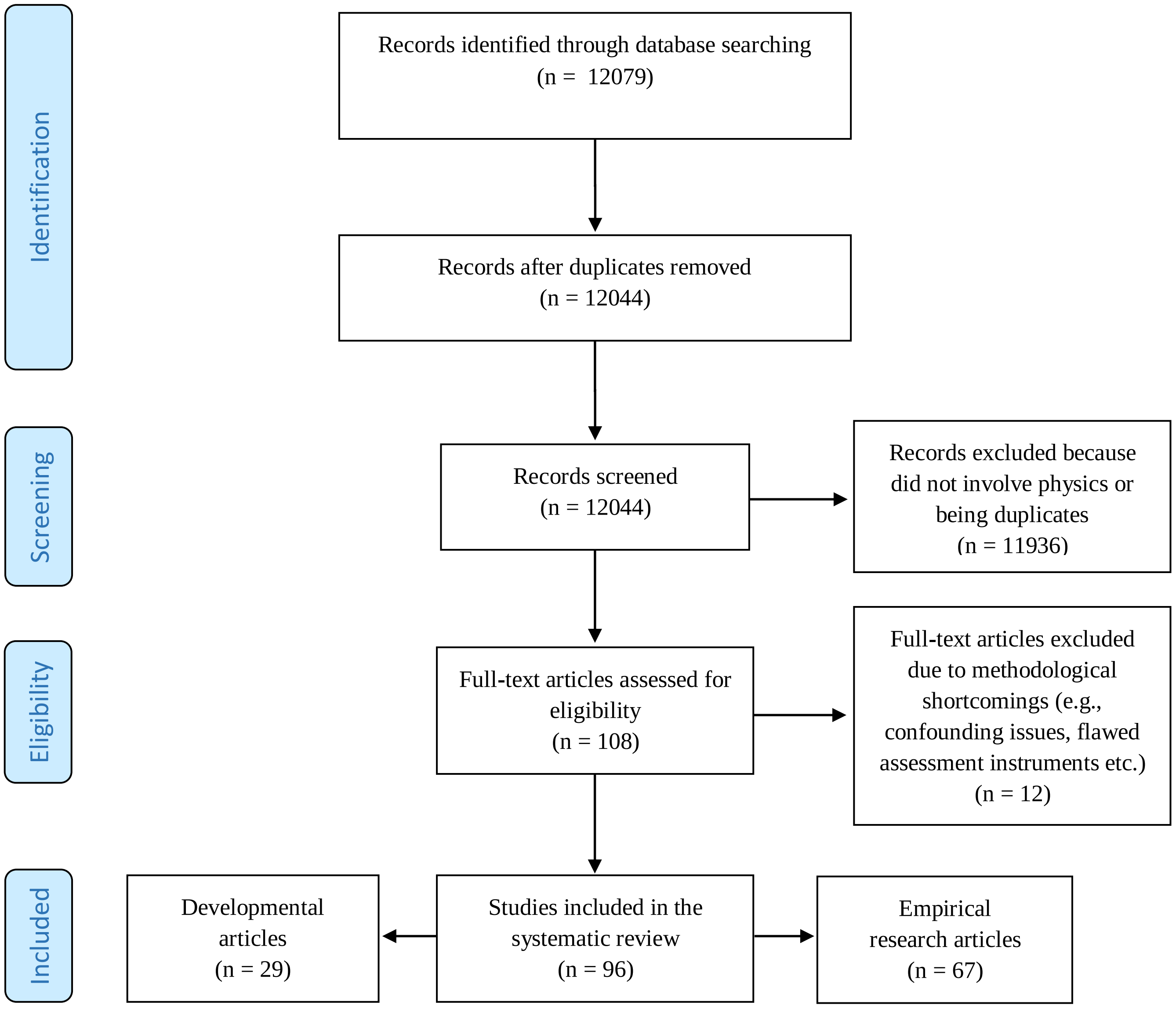

Figure 1. Article selection process based on Preferred Reporting Items for Systematic Reviews and Meta-Analyses (PRISMA) methodology [29].

Firstly, we searched for relevant papers in the Scopus and ERIC databases. The first search was conducted on January 1st, 2022 and it was supposed to include relevant articles published between January 1st, 2012 and January 1st, 2022. For purposes of updating the review, a second search was conducted on August 21st, 2023 and it included articles published between January 1st, 2022 and January 1st, 2023. Generally, we defined that a paper is relevant for our review if it met the criteria presented in table 2.

Table 2. Inclusion criteria.

| **Inclusion criteria** |
|---|
| Published between January 1st, 2012 and January 1st, 2023. |
| Peer reviewed journal article. |
| Related to using AR in teaching physics. |
| Written in English. |
| Original non-review article. |

Concretely, we applied the search terms "augmented reality" and "mixed reality" with the Boolean operator "OR" and specified that only peer-reviewed articles that had been published in English should be listed. Taking into account the rapid development of modern technologies and the requirement to keep our pedagogy up to date with these technological developments, we decided to include only articles that were published after January 1st, 2012. Additionally, we have excluded review articles, because they would not provide any new, original information about teaching ideas in AR-based learning of physics. By applying these search criteria to titles, abstracts and keywords of the indexed articles, we found 1160 articles in the ERIC database and 10919 articles in the Scopus database. In the next step, we had to identify those AR-articles that were related to learning about physics topics. To that end, we have operationally defined "physics" as "concepts and procedures that physicists develop or use in their study of natural phenomena, as described in physics textbooks at various levels of education". Both research and developmental (classroom physics) articles were considered relevant, because they both may be a source of ideas about AR-based learning about physics.

After thoroughly examining the titles, keywords, and abstracts of the 12079 search results, the first author of this review found that 143 of the results were related to learning about physics topics. These papers were considered potentially relevant for the systematic review. However, upon further examination of the 143 remaining articles, 35 of them were found to be duplicates and were removed, leaving 108 unique articles. In the second round of review, these 108 articles were again checked against the inclusion criteria and analyzed for major methodological and presentation deficiencies. This resulted in the removal of 12 more articles, eventually leaving 96 articles for a systematic review of augmented reality-based learning about physics. Some of the most common reasons for the removal of the 12 articles were as follows: the given educational technology does not constitute AR as defined in the XReality framework, measuring completely different constructs on a single scale (e.g., a single scale for measuring knowledge about solar systems and human body), using extremely small student samples in survey research (e.g., Rasch analysis based on a sample of approximately 20 students), confounding issues (e.g., AR use is potentially confounded by simultaneous use of other educational technologies), the targeted learning outcomes are not pure physics (e.g., knowledge about solar-power panel efficiency), and AR apps are not described at all in the given paper.

The whole initial review was done by the first author alone. Thereafter, for purposes of a reliability check, the second author independently reviewed a randomly selected subgroup of 25% of initially identified articles and assessed whether they met the article inclusion criteria. It was determined that the initial agreement between the first and second author was 93.7%. After further discussion, a complete agreement was reached. Both reviewers had more than 10 years of experience in teaching physics.

# 3. Findings and discussion

In this section, we present and discuss the findings for each of the research questions that were previously mentioned.

## 3.1. How are augmented reality technologies used for facilitating learning about individual physics topics?

Table 3 summarizes ideas about learning and teaching physics. Additionally, it provides an overview of the implemented research designs.

Table 3. A first summary of the reviewed literature with a focus on AR-based learning and implemented research designs; all 96 articles are included.

| Authors | Topics | Research design | AR-based learning |
|---|---|---|---|
| Abd Majid and Husain [30] | Rotation of the Earth and Moon phases | Developmental article | The AR application is intended for elementary school students. When they point their mobile phones at certain images in their textbooks, a 3D model showing the differences between day and night, or a 3D model of Moon phases is displayed on the phone screen. |
| Abdusselam and Karal [31] https://doi.org/10.1080/1475939X.2020.1766550 | Magnetic field | A quasi-experimental pretest/posttest study was conducted with two groups of students. In addition, the study included researcher observation and interviews. Data were analyzed quantitatively and qualitatively. | Students hold the mobile phone over the image marker and the 3D visualization of the magnetic field appears over the real objects. |
| Alhalabi *et al*. [32] https://doi.org/10.3390/educsci11110661 | Electrical circuits | Developmental article | A mobile application recognizes hand-drawn electrical circuit diagrams and generates circuit simulation results projected onto the input image. |
| Altmeyer *et al*. [33] https://doi.org/10.1111/bjet.12900 | Electrical circuits | An experimental pretest/posttest study was conducted with two groups of students. In addition, cognitive load and usability questionnaires were used. A quantitative data analysis was carried out. | Students run AR application and point the tablet at the printed markers. Augmented 2D image of the multimeter shows the measurement data on the display. |
| Alvarez-Marin *et al*. [34] https://doi.org/10.3390/electronics10111286 | Electrical circuits, Kirchhoff law | Survey research was conducted using multiple indicators for data collection. | Students used tablets to run AR application. By manipulating QR codes as targets, it is possible to interact with various electrical circuits components (batteries, light bulbs, resistors, and their voltages and resistances). |
| Aoki *et al*. [35] https://doi.org/10.1088/1361- | Kinematics | Developmental article | A 3D motion tracking system is built from a stereotype depth camera for |
| 6552/ab9213 | | | measurement of time and position data and a Raspberry Pi 4 for analysis. The trajectory of the object, velocity and acceleration vectors are displayed in AR by transmitting the 3D position data to the tablet device via Wi-fi. |
| Arici *et al.* [36] https://doi.org/10.1002/hbe2.310 | Solar system, Structure of the atom | Researchers used case study qualitative research design where they conducted interviews with students and teachers. | Users point tablets to printed images and they can observe virtual objects of celestial bodies in the real learning environment. |
| Baba *et al.* [37] https://doi.org/10.33200/ijcer.1040095 | Solar system and eclipses | A pretest-posttest quasi-experimental two-group design was used. The data collection instruments included the "Academic Achievement Test," the "21st-Century Skills Scale," and the "Augmented Reality Applications Attitude Scale." The data were subjected to quantitative analysis. | Students point tablets to the printed cards. The tablets' cameras recognize markers on the cards, allowing the precise overlay of 3D objects related to solar system and eclipses topics on top of the cards (e.g., 3D visualization of a planet). |
| Baran *et al*. [38] https://doi.org/10.1007/s10639-019-10001-9 | Electrical circuits | The researchers administered a pretest/posttest study with two groups of students. Observation notes and interviews with teachers were analyzed to collect data on the validity of the experimental design. Quantitative and qualitative data analyses were conducted. | Students used the AR application on tablets with printed markers to trigger 3D digital visualizations of electrical circuit elements. |
| Bodensiek *et al*. [8] https://doi.org/10.1119/1.5095388 | Magnetism | Developmental article | Observing the experimental setup in real time through the smartglasses, the user can see the visualization of the magnetic field as a vector diagram. In addition, the numerical values of the variables in the experiment and relevant formulas are displayed. |
| *Cai et al.* [39] https://doi.org/10.3390/su14116783 | Acoustics | A single group mixed-methods research was implemented. The data were collected using Conceptions of Learning Science and Scientific Epistemic Beliefs questionnaires, and semi-structured interviews. The researchers used quantitative and qualitative techniques to analyze the collected data. | During the learning about the Doppler effect, students worked collaboratively, with one student holding a tablet and another holding marker cards. An augmented reality image (i.e., wave fronts) was generated and displayed on top of the marker cards. By moving the marker cards students could simulate the motion of a wave source relative to an observer which affected the appearance of the wave fronts. |
| Cai et al. [40] https://doi.org/10.1007/s10956-022-09991-y | Lever principle | A quasi-experimental two-group pretest/posttest design was implemented. Students completed questionnaires that evaluated their knowledge, self-efficacy, flow, and cognitive load. Furthermore, researchers conducted interviews. The collected data was subjected to a thorough analysis using both quantitative and qualitative data analysis techniques. | Students directed tablets towards recognition markers, prompting the mobile AR application to run a lever principle virtual experiment within the real world scenery. Students could adjust the masses of the weights and place them on the lever. Furthermore, students wore EEG sensors on their heads which measured their attention levels, while the AR application delivered real-time visual feedback corresponding to their current level of attention. |
| Cai *et al*. [41] | Convex lens | An experimental pretest/posttest study was conducted with two groups of students. Additionally, the experimental group of students completed an attitude questionnaire after the teaching intervention. Quantitative analysis was used. | The laptop camera is pointed at the printed markers and the 3D objects of the convex lens, the candle and the optical image are displayed on the screen in AR environment. |
| Cai *et al*. [9] https://doi.org/10.1080/10494820.2016.1181094 | Biot-Savart Law | In a quasi-experimental study, students were administered a pretest, a posttest, a delayed posttest, and an attitude questionnaire, while several randomly selected students were also interviewed. Quantitative and qualitative analyses were conducted. | Students wave their hands to trigger a virtual 3D magnet model and the simulated magnetic field using a motion-sensing environment. |
| Cai *et al*. [42] https://doi.org/10.1111/bjet.13020 | Photoelectric effect | A quasi-experimental study was conducted using a pretest/posttest questionnaire. Quantitative analysis of data was used. | Tablets are pointed at printed markers and in the mobile AR application, students can change variables that affect the virtual 3D environment (e.g., resistance, current). |
| Çakıroğlu *et al.* [43] | Lunar and solar eclipse | A case study was conducted. Concretely, to assess the impact of | During the process of constructing the concept map, students run the app and turn |
| https://doi.org/10.1186/s41039-022-00191-1 | | this application on students' concept map creation skills and their perceptions of their experiences, observations were made using video recordings and open-ended question worksheets. The researchers utilized qualitative and quantitative techniques to analyze the collected data. | their phones towards the QR code located beside the relevant astronomical concept. Thus, an augmented reality image is generated, presenting the 3D shapes related to each astronomical concept along with their connections to other concepts. |
| Chandrakar *et al.* [44] https://doi.org/10.1119/10.0002739 | Projectile motion | Developmental article | In this AR based game students use the smartphone camera as AR smartglasses. They define a real object as a target and predict the initial velocity and angle of the throw of a virtual projectile. As they shoot the virtual projectile, they can observe its trajectory. If the target is missed, the application gives an indication of how to change the parameters. |
| Chang and Hwang [45] https://doi.org/10.1016/j.compedu.2018.06.007 | Magnetic force and motor | An experimental pretest/posttest study was conducted with two groups of students. In addition, five learners from each group were interviewed. Quantitative and qualitative data analyses were performed. | Students point the tablets at the real objects and AR application is displaying information and instructions on how to use objects related to the construction of an electric motor. Additionally, the AR system can provide students with a 3D visualization of magnetic force and magnetic flow. |
| Chen *et al.* [46] https://www.jstor.org/stable/48660130 | Motion of planets | The researchers employed a quasi-experimental two-group pretest/posttest research design. They used questionnaires to gather data on students' knowledge, motivation, and flow. Additionally, students from the experimental group were interviewed. Both qualitative and quantitative techniques were applied to analyze data. | Students aim their tablets at image markers, which results in 3D content related to planet motion being shown on top of real-world scenery. |
| Coşkun and Koç [47] https://doi.org/10.52963/PERR_Biruni_V10.N2.21 | Solar system and Universe | Pre-test-post-test control group quasi-experimental design was used. Data was collected through the "Solar System and Beyond Success Test"; "Science Learning Anxiety Scale", "Astronomy Attitude Scale" and "Students' Motivation Scale for Science Learning" (SMSS). | First students observed sky trough he mobile application installed on tablet computer. Additionally, they point tablets to printed image markers that represents different planets. |
| Costa *et al.* [48] https://doi.org/10.3390/mti5120082 | Celestial bodies and planetary systems | Mixed methodology with a qualitative and interpretative approach that included online participant observation and questionnaire. Group of teachers was tested with the aim of assessing perceptions about used system. | Game application is running at GPS compatible smartphones and users are visiting specific outdoor locations. Virtual objects that represent celestial bodies appear on the top of real world scenery and are visible at the screen of their devices. During the game users need to move in real world and capture virtual objects, which are stars, planets and other celestial bodies. |
| Daineko *et al.* [49] https://doi.org/10.1002/cae.22297 | Thermodynamics, Molecular physics, Mechanics, Vibrations and waves | The researchers used three surveys to assess student attitudes toward past use, ease of use, and satisfaction. Teachers were also interviewed. Quantitative and qualitative analyses were conducted. | Students point their mobile phone at printed markers and use the AR application to learn about various physics topics (e.g., Atwood machine, Maxwell pendulum). |
| Daineko *et al.* [50] https://doi.org/10.3991/ijim.v14i13.13475 | Electrical circuits, kinematics, moment of inertia | Developmental article | Students launch the AR application and then select an experiment from the menu. The corresponding theory and instructions on how to perform the experiment in conjunction with the AR application are displayed. |
| Demircioglu *et al.* [51] https://doi.org/10.31014/aior.1993.05.02.464 | Astronomy | A quasi-experimental design involving two experimental groups and one control group was employed. Data were gathered using pre- and post academic achievement tests, as well as the Motivated Strategies for Learning Questionnaire (MSLQ). | Students used tablet computers through free applications such as i-solar system, Aurasma, Junaio, Sky view Free, Augment and Star Chart. Videos, Simulations, and 3D visuals about astronomy were used as "overlays" during the activities. "Trigger images" were photographs, colored areas and pictures in the students' textbooks and worksheets. |
| Dilek and Erol [52] https://doi.org/10.1088/1361-6552/aaadd9 | Kinematics | Developmental article | The mobile phone is fixed to a moving cart to record the horizontal position of the cart, and the application creates real-time position-time graphs from the measured data. |
| Donhauser *et al.* [53] https://doi.org/10.1119/10.0001848 | Magnetism | Developmental article | Through the smartglasses, the user can virtually visualize the model of the magnetic field lines directly around the horseshoe magnet, along with the displayed formula and vector directions. |
| Düzyol *et al.* [54] https://doi.org/10.31681/jetol.976885 | Astronomy | The researchers employed a pretest-posttest control group design along with semi-structured interviews. Both, quantitative and qualitative data analyses were utilized. | Students aim tablets to the space-themed card set and 3D astronomical objects are displayed on the screen on top of the corresponding cards. |
| Echeverría *et al.* [55] https://doi.org/10.1016/j.chb.2012.01.027 | Coulomb's law | A pretest-posttest design was conducted. The researchers conducted five sessions in which students were randomly assigned to three groups of three. Data were analyzed quantitatively. | Tablets are pointed to markers on the table. Students vary the position of the 3D game characters shown on the tablet displays by changing the distance between the tablet and the markers. By changing the electrical properties (on the tablet screen) and the positions of the game character positions, they learn about electrical forces. |
| Enyedy *et al.* [56] https://doi.org/10.1007/S11412-012-9150-3 | Force, friction and two-dimensional motion | The researchers conducted a pretest/posttest case study. They used interviews, questionnaires, videotapes, and problem-solving activities. Quantitative and qualitative analyses were conducted. | Students wear special printed symbols and place others on the floor. Their movements are tracked and AR computer application recognizes the symbols using cameras mounted on the ceiling. As a result, 3D objects (e.g., a ball) are displayed on the screen together with real objects. |
| Estrada *et. al.* [57] https://doi.org/10.3390/info13070336 | Electrical circuits | Developmental article | The smartphone AR application utilizes Deep Learning (DL) algorithms to achieve real-time, highly precise detection and recognition of specific electrical equipment within the laboratory. Following recognition, a 3D model is overlaid onto the physical object in the app. Furthermore, the app enhances the user experience by presenting additional information about the equipment in the form of holograms. |
| Faridi *et al*. [58] https://doi.org/10.1002/cae.22342 | Electromagnetism | A pretest-posttest design was conducted with two groups of students. In addition, the researchers interviewed students in the experimental group. Quantitative data analysis was conducted. | Students point their mobile phones at printed markers and AR application shows 3D objects (magnets, solenoid, multimeter, etc.) over a real scenery. |
| Fidan and Tuncel [59] https://doi.org/10.1016/j.compedu.2019.103635 | Weight, mass, pressure, force, work, energy | A quasi-experimental design was conducted with two experimental groups and a control group. In addition, the researchers used semi-structured interviews. Qualitative and quantitative data analyses were conducted. | Students used AR application on tablets with the special tracker marker cards to display the 3D virtual objects in the real world (e.g., dynamometer). |
| H. Y. Wang *et al.* [60] https://doi.org/10.1007/s10956-014-9494-8 | Elastic collision | The students were randomly assigned into groups of two and distributed between an experimental and a traditional group. Video recordings of the students' inquiry and discussion process were analyzed using content and lag sequential analysis. | Students run the mobile AR application and point their phones at the printed markers. On the screen they see virtual 3D cubes colliding elastically with each other. |
| Huang and Lin [61] https://doi.org/10.1108/LHT-01-2017-0023 | Stars and constellations in the sky | An experimental design was conducted with two groups of students. A pretest and posttest on knowledge and a questionnaire on self-efficacy were administered. Quantitative data analysis was used. | Students point their phones at the local sky and AR application displays images of stars and constellations based on current location, elevation angle, and local time. |
| Hyder *et al.* [62] https://doi.org/10.14569/IJACSA.2021.0120284 | Particle physics | Scientists used qualitative evaluation and quantitative one group method based on a survey. Participants were college students. | Kinect motion tracking sensors and infrared-based camera were used to track the body motion. The developed environment is projected on the floor providing real-time interactive learning experience. System simulates LHC tunnel to visualize the generation of new particles |
| | | | after the collision of protons. Students can play a football game in which the balls are representing protons. |
| Ibáñez *et al.* [63] https://doi.org/10.1016/j.compedu.2013.09.004 | Coulomb's law, electrical current, electric field, Ohm's law, Lorentz's law | A pretest/posttest followed by a Flow State Scale questionnaire was administered to all students in the experimental group. In addition, the experimental group answered an open-ended questionnaire about their attitudes. The researchers conducted both quantitative and qualitative analyses of the data. | Students aimed tablets at marker-based 3D shapes that mimicked circuit elements. AR application enables 3D visualization of topics in electromagnetism. |
| Ibanez *et al.* [64] https://doi.org/10.1109/TE.2014.2379712 | Electrical current and resistance | The researcher administered Pintrich's Motivated Strategies for Learning Questionnaire (MSLQ) prior to the activity. After each experimental activity, students were asked multiple-choice questions, and a motivation survey was administered at the end of the activities. In addition, ten students' activities were videotaped. Quantitative and qualitative data analyses were performed. | Students run mobile application and point the tablet at the printed markers to activate AR. To build electrical circuits, students manipulated 3D shapes on their mobile screen. |
| Jiang *et al.* [65] https://doi.org/10.1177/07356331211038764 | Thermodynamics | Researchers observed students engaging in science experiments in pairs. They collected data through semi-structured interviews, logs from the augmented reality (AR) application, and laboratory reports. The gathered data underwent both quantitative and qualitative analysis. | Each pair of students had a smartphone that they directed towards the experimental setup (paper and jar). Students used an AR application named "Infrared explorer" to learn about radiation, convection and radiation (i.e., explore the temperature of the paper when the jar was placed at different places). The AR app allowed students to see a thermal version of a laboratory experiment using the infrared camera attached to a mobile device. |
| Kapp *et. al.*[66] https://doi.org/10.3390/s22010256 | Electrical circuits | Developmental article | While wearing Microsoft HoloLens 2 equipped with an AR application, students assemble electrical circuits using provided sensor boxes containing resistors, power supplies, and cables. The AR application displays real-time measurement data directly above the respective sensor box positions. |
| Kasinathan *et al.* [67] https://doi.org/10.30880/ijie.2018.10.06.021 | Solar system | Developmental article | Users point their mobile phone at one of the 10 target images in a traditional book and AR application displays the 3D model of the image. The user can interact with and explore the 3D model. The images have a pop-up display with a simple image description complemented by a spoken audio recording. |
| Keifert *et al.* [68] https://doi.org/10.1080/09500693.2020.1851423 | Water states of matter | Extended sequential analysis of videotaped students' activity. Additionally, researchers conducted student post-interviews. | Computer based body-tracking technology follows students' movement and communicates that information to a simulation. The environment combines that data with a computational simulation that models how water particles move in different states of matter. |
| Kirikkaya and Başgül [69] https://doi.org/10.33225/jbse/19.18.362 | Solar system and celestial bodies | An experimental pretest/posttest study was conducted with four groups of students. The researchers used quantitative data analysis. | Students point their mobile device at the sky and AR application is showing images of 3D visualizations of celestial bodies. |
| Lauer *et al*. [70] https://doi.org/10.1119/10.0002078 | Electrical circuits | Developmental article | Either with smartglasses firmly mounted on the user's head or with a tablet attached to a stand, users can observe AR circuit symbols displayed over the real individual circuit components. As their hands are free, they can assemble real circuits and the virtual circuit diagram adjusts accordingly. |
| Lin *et al*. [71] https://doi.org/10.1016/j.compedu.2013.05.011 | Elastic collision | Students were divided into groups of two and assigned to one experimental and one traditional group. In addition, participants' learning activities were video and audio recorded for later | Students point their mobile device at the printed markers and the augmented 3D objects are superimposed on the real environment. Elastic collisions are simulated by manipulating 3D objects in |
| | | analysis. The researchers used quantitative and qualitative data analysis. | the AR application. |
| Lincoln [72] https://doi.org/10.1119/1.5055344 | Moon | Developmental article | The AR application is activated when the mobile device's camera points to different areas of the lunar model and geological and historical information appears in the form of virtual text or 3D animations. |
| Lindgren *et al.* [10] https://doi.org/10.1016/j.compedu.2016.01.001 | Gravity and planetary motion | The researchers conducted pre- and post- surveys for one group of participants. Quantitative analysis was used. | In the room-sized AR simulation, space images are projected onto the floor. Students impersonate meteors and their movement is tracked using laser scanning technology. |
| Lindner *et al.* [73] https://doi.org/10.1016/j.actaastro.2019.05.025 | Astronomy, Gravitational force | Developmental article | The 2D image of Earth is a marker for AR, which turns it into a rotating 3D Earth and places Moon right in front of the smartphone's camera. The app displays the time it takes Moon to revolve around Earth, depending on the distance between the two objects. |
| Liou *et al.* [74] | Moon phases and trajectory | The researchers conducted a quasi-experimental design with pretest/posttest, task tests, technology acceptance questionnaire, and interview. Students were divided into teams of two or three in the control and experimental groups. Quantitative and qualitative data analyses were used. | Students point tablets at the sky to find Moon. AR application is showing the image of Moon overlaid on the real environment. |
| Liu *et al.* [75] https://doi.org/10.1111/jcal.12513 | Magnetic field | Experiment adopted randomized three groups Pretest-Posttest-Control design where participants were assessed by knowledge quiz, cognitive load scale and perceptions survey. Finally, semi-structured interviews were conducted with three students. | Students point tablet with running AR application to the real magnet that served as AR marker to observe augmented field lines. Furthermore, image markers were used for visualization of the geomagnetic field. |
| Mahanan *et al.* [76] https://doi.org/10.3991/ijim.v15i23.27343 | States of matter | Developmental article | An augmented reality application on a mobile device is used to visualize particles (number, size, arrangement and distance between them) as the liquid turns into a solid in the student's project activity (secondary school): Making homemade ice cream. |
| Majid and Majid [77] https://doi.org/10.18517/ijaseit.8.4-2.6801 | Atomic models | A pretest/posttest design was conducted with one group. The researchers used quantitative data analysis. | Students bring their smartphones over the printed markers and can see a 3D model of atoms, images, and videos through the AR application. |
| Matcha and Rambli [78] https://doi.org/10.11113/jt.v78.6941 | Electrical current and resistance | A case study was conducted and data was collected by using video observation and questionnaire. Qualitative and quantitative analysis was used. | The computer camera is pointed at the printed markers and a 3D visualization of the electrical circuit components appears on the computer screen. In addition, the AR ammeter can be used for measurements. |
| Matsutomo *et al.* [79] https://doi.org/10.1109/TMAG.2013.2240672 | Magnetism | Developmental article | Web camera captures images of mocks (representing a bar magnet and a piece of iron) placed on a background monitor. As a result, a 2D magnetic field is immediately displayed on the monitor. |
| Matsutomo *et al.* [80] https://doi.org/10.1109/TMAG.2017.2665563 | Magnetism | Developmental article | Users can easily observe the magnetic field generated by one or more sources in augmented 3D space using smartglasses. They can also move the sources and observe how the magnetic field distribution changes in real time. |
| Montoya *et al.* [81] https://doi.org/10.12973/eurasia.2017.00617a | The structure of the atom Electrical circuits | A quasi-experimental study was conducted with two groups of students using a pretest/posttest and an attitude survey about AR Learning. Data were analyzed quantitatively. | Students point the mobile device at the printed marker and in the AR application 3D models of the selected objects/phenomena are displayed, complemented by audio and text descriptions. |
| Moriello *et. al.* [82] 10.1109/ACCESS.2022.3175869 | Electromagnetic radiation - gamma radiation | Developmental article | Students point their phone's camera at suitable targets (representing fake detector and radioactive source) and representation of the experiment equipment is displayed in the AR application. By manipulating |
| | | | with target detector, students can set distance and inclination with respect to the target source. Sensors measure these values and send them to the real remote laboratory to move and arrange real source and detector. Students can retrieve and view the corresponding gamma-ray energy spectrum on their smartphones. |
| Mukhtarkyzy *et al.* [83] | Electrical circuits | A quasi-experimental design with two experimental and two control groups was applied. Pre-test survey and post-test quiz was administered to all students. Furthermore, students from the experimental group were administered an additional post-test survey. Quantitative techniqueswere used to analyze the collected data. | The application displays 3D models of electrical circuits components printed in the book, and the students can interact with the contents of the textbook by pointing their tablet device's camera at the printed picture. The application also includes demonstration exercises that allow students to connect equipment and complete electrical circuits on their tablets. |
| Oh *et al.* [84] https://doi.org/10.1109/TLT.2017.2750673 | Refraction of light | Researchers distributed dyads of students to experimental and control group. Pretest, posttest, attitude survey and semi-structured interview were used to collect data. Quantitative and qualitative data analyses were performed. | In a room-sized AR environment, students learned about light refraction using smartglasses and body gestures. The system combines 3D visualizations projected globally onto the floor with transparent graphics projected locally from the AR glasses. |
| Önal and Önal [85] https://doi.org/10.1007/s10639-021-10474-7 | Solar system and Universe | Researchers used mixed method pretest-posttest two random group experimental design. Achievement and interests in both groups were assessed by tests while semi-structured interview was administered only to experimental group. | Students points mobile phones or tablets to printed cards and virtual 3D astronomical objects appear on the mobile phone screen. |
| Özüağ *et al.* [86] https://doi.org/10.18178/ijeetc.8.1.9-13 | Electrical circuits | Developmental article | The AR interactive application simulates electronic circuits on the display of an Android smartphone, using circuit diagrams drawn on paper as markers. The value on the potentiometer can be changed on the smartphone screen, changing the current flow in the circuit. This is indicated by yellow dots whose size varies accordingly. |
| Özüağ *et al.* [87] https://doi.org/10.1142/S0218126620500966 | Electrical circuits | Developmental article | AR application simulates two AC electronic circuits when users point their smartphones at the printed markers. In addition, the circuit elements can be easily controlled via the touchscreen. |
| Phon *et al.* [88] https://doi.org/10.1166/asl.2015.6307 | Trajectory of Earth, Moon and Sun | The researchers conducted a pretest/posttest study with one group. A quantitative analysis was conducted. | Students held printed markers in front of the desktop computer camera and a virtual 3D object from astronomy appeared on the digital screen. |
| Restivo *et al.* [89] https://doi.org/10.3991/ijoe.v10i6.4030 | Electrical circuits (DC) | An experiment was conducted with two groups. The researchers used surveys, field notes from an outside observer, audio recordings, screen capture and students' wiring activity. Qualitative and quantitative analyses were used. | Web camera is pointed at the printed markers of the DC circuit. The computer AR application displays the 3D components of the DC circuit above the markers. |
| Reyes-Aviles and Aviles-Cruz [90] https://doi.org/10.1002/cae.21912 | Electrical circuits (DC) | Researchers carried out a qualitative survey study with two groups of students. | A student photographs a typical resistor circuit, and the markerless mobile AR application can detect and classify resistors and generate virtual 3D objects of circuit components. |
| Rosi *et al.* [91] https://doi.org/10.1088/1361-6552/abd5a2 | Kinematics | Developmental article | A 3D motion tracking system set from headset and its controllers sends the signal to sensors allowing the computer to track a student's position and movements. In addition, a stereo camera is connected so that the captured and calculated data (graph, velocity and acceleration vectors...) are displayed in the headset in real-time. |
| S. A. Yoon and Wang [92] https://doi.org/10.1007/s11528-013-0720-7 | Magnets and magnetic fields | A quasi-experimental pretest/posttest study was conducted. In addition, the researchers used interviews to understand the impact of AR on learning. Quantitative and qualitative | Students hold a magnet in front of the computer camera and AR magnetic field lines appear around the magnet on the computer screen. |
| | | analyses were conducted. | |
| S. A. Yoon *et al.* [93] https://doi.org/10.1007/s11412-012-9156-x | Electrical circuits and conductivity | The researchers combined a quasi-experimental design with a qualitative design and included four groups of students. Surveys, student response forms, interviews, observation notes, and video footage of student interaction with the devices were used. Quantitative and qualitative analyses were conducted. | Students touch metal spheres and hold their hands to form an electrical circuit. When the circuit is completed, projected AR animations visualize the flow of electrons through spheres and the students' bodies. |
| S. A. Yoon *et al.* [94] | Bernoulli principle | The researchers conducted a quasi-experimental design with two groups (pretest/posttest) followed by interviews. Quantitative and qualitative analyses were conducted. | Students place a real ping pong ball over a real air blower. On the screen they see a digital 3D ball and AR arrows representing the air movement around the ball. |
| Sahin and Yilmaz [95] https://doi.org/10.1016/j.compedu.2019.103710 | Celestial bodies, solar system and space research | A quasi-experimental study was conducted with two groups. In addition, the experimental groups answered an attitude questionnaire. The researchers used a quantitative analysis of the data. | During the lectures teachers used desktop computer with projector to show students virtual celestial objects from the AR based activity booklet. |
| Sanderasagran *et al.* [96] | Fluid flow | Developmental article | Users point their mobile devices at the printed marker and AR application displays the two-dimensional fluid flow behavior for three simple geometries. |
| Selek and Kiymaz [97] https://doi.org/10.1002/cae.22204 | Electrical circuits (current, voltage) | Experimental study included six control and six test groups. Researchers used small written or oral exams. The applied lessons grades and the academic averages at the end of first and second semesters of 48 students were taken into consideration. | Students point mobile phones to image markers and virtual electric circuits with elements are shown on the screen. |
| Seo and So [98] https://www.jstor.org/stable/48695978 | Conductor resistance | A quasi-experimental study involving two groups of students was conducted. The cognitive learning impact was measured through a multiple-choice test. | Students stand in front of a mirror-like AR system that is using a motion sensor to recognize their images in real-time and perceive gestures using balance and eyesight. Students' hand movements (horizontal and vertical) change the length of the displayed electrical conductor, bulb brightness and current intensity. Here haptical elements from the real-world context are combined with certain virtual objects (e.g., appearance of the electrical conductor). |
| Sonntag and Bodensiek [99] https://doi.org/10.1103/PhysRevPhysEducRes.18.023101 | Electrical circuits | The researchers employed a pretest/posttest true experimental design with four groups of students. Eyetracking analyses, self-reported measures, and questionnaires as assessment tools were used. Both quantitative and qualitative data analysis techniques were applied to comprehensively examine the collected data. | Students focused their gaze on an electrical circuit, using a head-mounted display (MR-HMD) to observe augmented 3D visualizations (i.e. electric current, magnetic fields). |
| Stolzenberger *et al.* [100] https://iopscience.iop.org/article/10.1088/1361-6552/ac60ae | Electrical circuits | Developmental article | The AR application for mobile devices enhances the user's understanding of electric current and electric potential by virtually superimposing moving electrons (depicted as white spheres) and electric potential (indicated by color) onto an experimental kit. Each circuit component has a printed marker taped to it which allows the AR app to identify the component. |
| Strzys *et al.* [101] https://doi.org/10.1088/1361-6404/aaa8fb | Heat conduction in metals | The researchers used a quasi-experimental design with two groups. Data were collected using an introductory test, a pretest, and a posttest. A quantitative data analysis was performed. | Experimental laboratory setup was connected with AR application run by smartglasses. Wearing the smartglasses students can see augmented color image of temperature values on real metal rod and data graphs plotted in real-time. |
| Sung *et al.* [102] https://doi.org/10.3390/app9194019 | Free fall | Developmental article | Users watch the Kinect video stream of a selected surface in combination with the AR simulation of a free fall of a soft body. |
| Suprapto *et al.* [103] | Atomic | A single-group case study was | Students point their smartphone at the |
| https://doi.org/10.3991/ijet.v15i10.12703 | models | conducted that included a posttest and a questionnaire on augmented reality media use. | printed marker and AR application presents a 3D model of the atom. |
| Suprapto *et al.* [104] http://dx.doi.org/10.3926/jotse.1167 | Planetary motion, Gravitational force, Kepler's laws | Researchers used one-group pretest-posttest design. Multiple-choice questions that assessed students' achievement were the same in the pretest and posttest. | Students point mobile phones to the printed 2D pictures and inside AR application, multiple objects augmented to the real picture can be seen. |
| Talan *et al.* [105] http://dx.doi.org/10.21891/jeseh.1193695 | Solar system | A pre-test/post-test two-group quasi-experimental design was implemented. Furthermore, students from the experimental group took part in semi-structured interviews. The researchers employed both qualitative and quantitative techniques for data analysis. | Students direct cameras from tablets, phones or computers to the flash cards in order to activate corresponding 3D content. In other words, AR is used to enhance the didactic potential of flash cards within the context of astronomy. |
| Tarng *et al.* [106] https://doi.org/10.1155/2018/5950732 | Sun path concepts | A pretest/posttest design was conducted with two groups. In addition, interviews were conducted with the students in the experimental group and the teacher to determine the validity of the results. Quantitative and qualitative data analyses were conducted. | The student points his mobile device at the Sun and observes and notes its orientation and elevation. In addition, AR application displays a virtual Sun. |
| Teichrew and Erb [107] https://doi.org/10.1088/1361-6552/abb5b9 | Freebody diagrams, electrical circuits and optics | Developmental article | Users point their mobile phones at the real experimental setups and AR application visualizes physical quantities that cannot be seen in the experiments, e.g., forces on an object on the inclined plane, potential differences in electrical circuits and light rays on plane mirrors. |
| Thees *et al.* [108] https://doi.org/10.1016/j.chb.2020.106316 | Heat conduction in metals | Researchers implemented a quasi-experimental pretest/posttest design with two groups of students. Quantitative data analysis was conducted. | Experimental laboratory setup was connected with AR application run by smartglasses. Wearing the smartglasses students can see augmented color image of temperature values on real metallic rod and real-time data graphs. |
| Tian *et al.* [109] https://doi.org/10.3991/ijim.v8i1.3457 | Lunar phase | A case study that included data collection through a questionnaire and interviews was conducted. Quantitative and qualitative data analyses were performed. | Students point smartphones at the sky to find the Moon. AR application shows the images of the Moon phase, horizon, latitude and longitude overlaid with the real environment. |
| Tomara and Gouscos [110] https://doi.org/10.1177/0735633119854023 | Coulomb's law | Researchers conducted a one group case study which included a survey and Likert-type scale questionnaire. Quantitative analysis was used. | Students point their smartphones or tablets at printed markers and in the AR application 3D models of point charges are visualized along with force vectors. Additionally, information about charges and force is displayed. |
| Tscholl and Lindgren [111] https://doi.org/10.1002/sce.21228 | Force and motion in a planet and asteroid context | A mixed methods study was conducted. Observational data (written notes) and video and audio recordings of interactions were used. The material was transcribed and a coding scheme was implemented. The researchers used correlation analyses of the data. | In the room-sized AR simulation, space images are projected onto the floor. Students impersonate meteors and their movement is tracked using laser scanning technology. |
| Urbano *et al.* [112] https://doi.org/10.1002/cae.22306 | Electrical circuits (DC) | Researchers implemented a one group pretest/posttest design. Qualitative analysis was conducted. | A computer AR application uses a camera that points at printed markers, recognizes them, and overlays a real circuit diagram with 3D electrical circuit components on the screen. |
| Vidak *et al.* [113] https://doi.org/10.1088/1361-6552/ac21a3 | Gravitational force | Developmental article | When students view the markers through an AR mobile application, they see two 3D spheres above the papers. By moving the printed markers, students can change the distance between the virtual spheres, and on the screen, they can directly change the mass of the spheres. The distance between the centers of the spheres and the magnitude of the gravitational force between them are displayed on the screen, with vectors representing the directions of the gravitational forces acting on the spheres. |
| Vidak *et al.* [114] https://doi.org/10.1119/5.0037214 | Rolling motion | Developmental article | The AR application for smartphones uses the printed QR code as a target to display animations of the rolling motion of four physical bodies down an incline on top of real-world scenery. It also visualizes static friction force and velocity vectors at the point where each object makes contact with the surface. |
| Villanueva *et al.* [115] https://doi.org/10.1186/s41239-021-00268-9 | Electrical circuits | Researchers conducted Q-matrix assessment design where electrical circuitry micro skills were investigated in partial AR and full AR group of students. Multiple-choice questionnaire was used to asses students' knowledge and think aloud protocol was implemented to gain insight into students' learning process. | Mobile phone is attached to the stand and students deliver electrical circuit components under the mobile phone camera to get augmented information (animations, video tutorials, 3D objects, and components info). |
| Volioti *et. al.* [116] https://doi.org/10.3390/info13070336 | Measurements in physics | Developmental article | There are three distinct AR applications, each designed for a particular school grade level. Within each application, there are six sections, each detailing a physics experiment. To initiate each experiment, students begin by scanning a QR code found in their textbooks. This action triggers the appearance of virtual objects above the textbook page, representing the necessary materials for conducting the specific experiment. |
| Wang [117] https://doi.org/10.1080/21548455.2022.2072015 | Magnetic field | A quasi-experimental pre-test/post-test design was used, involving two experimental groups and one control group. The researcher employed the learning satisfaction scale, analyzed classroom interactions, and conducted interviews with five students from each group after the experiment. Both quantitative and qualitative data analysis techniqueswere utilized. | Students direct their mobile devices towards physical bar and shoe magnets in the real world and an augmented reality (AR) application is utilized to showcase virtual objects such as paper clips and field lines, providing a visual representation of magnetic phenomena. |
| Wildan *et al.* [118] https://doi.org/10.1119/5.0037354 | Electrical discharge | Developmental article | When the mobile device is turned towards the markers placed on the table, a virtual van de Graaff generator together with a discharge wand is shown on top of the real-world table. The user may modify the motor speed of the generator, the distance between the generator and spherical wand, and investigate how much time is needed for the generator to charge up, for a fixed motor speed. |
| Xiao *et al.* [119] https://doi.org/10.4018/IJDET.2020010102 | Characteristics of eight planets in solar system | A single-group study was conducted that included a basic information survey, a learning attitude survey, and an EEG measurement. Quantitative analysis was performed. | Tablet AR application was triggered by custom symbol cards. Augmented 3D objects showed eight planets and provided information about them. |
| Y. H. Wang [120] | Building of electrical scientific toy | A pretest/posttest design with two groups was used. In addition, teacher and several students from each group were interviewed. Qualitative and quantitative data analyses were conducted. | Students point the tablets at the real objects and AR application is displaying information and instructions on how to use the objects to facilitate the construction of an electrical scientific toy. |
| Yau *et al.* [121] https://doi.org/10.1088/1361-6552/ab7ae4 | Discharge of plasma | A case study was conducted with pretest/posttest surveys. The researchers used quantitative data analysis. | Students move their mobile devices towards printed markers placed on real world objects. AR recognizes the markers and displays the 3D objects from the plasma discharge experimental setup. |
| Zhang *et al.* [122] https://doi.org/10.1016/j.compedu.2014.01.003 | Stars and constellations in the sky | A quasi-experimental study was conducted with 4 groups. All groups were administered a pretest/posttest and a flow experience questionnaire after the activity. In addition, the same flow experience questionnaire was administered to the students thirty days after the activity. A quantitative data analysis was performed. | Students use mobile devices with digital compass and G-sensor. They point the devices at the sky and AR application automatically displays images of constellations. |

From table 3 we can conclude that a whole range of AR features can be used to facilitate learning of various physics topics. By analyzing the different learning activities, we could identify the following ways in which AR may facilitate physics learning:

1) Facilitating learning by overlaying abstract representations on top of real-world scenery (e.g., vectors, field lines, streamlines, rays)
2) Facilitating learning by visualizing relevant microscopic processes (e.g., flow of charges through a circuit)
3) Facilitating learning by a systematic, organized presentation of relevant factual information on top of real-world scenery (e.g., instructions on how to assemble an electric motor or how to use experimental equipment)
4) Facilitating learning by including haptic experiences and allowing for analyzing cause and effect relationships (e.g., verifying Newton's law of gravitation)
5) Facilitating learning by enriching the real-world scenery with data obtained from certain sensors (e.g., GPS data which helps to identify constellations or temperature sensor data that allow us to differentiate the temperature of real-world objects, based on the overlaid color-image)
6) Facilitating learning by improving the quality of learning materials (e.g., allowing the possibility to generate 3D visualizations that extend the textual presentation)

Mostly, AR applications facilitate learning by providing 3D visualizations. Thereby, AR is used to combine the authentic representation of the given physical phenomenon with an auxiliary visual representation that is part of the augmentation feature (e.g., real-world magnet and virtual field lines as an auxiliary visual representation). This is consistent with Ainsworth's complementary function of multiple representations [20]. In addition, there are also cases where the auxiliary visual representation (which is part of the augmentation) has the function of preventing the development of misinterpretations which corresponds to Ainsworth's notion of multiple representations that work together to limit the room for interpretations (e.g., the student's movement is tracked with a motion sensor and the corresponding movement is displayed on the screen, which can prevent the development of the graph as an image of the motion misinterpretation). Sometimes combining multiple representations may induce cognitive overload. This happens particularly often when some of the representations include many features irrelevant to achieving the learning objectives. For example, this may be the case for the activity that includes playing football with balls that represent protons [62]. In order to be in position to more objectively evaluate the effectiveness of learning activities, it would be useful if future AR-studies would more often report the different types of cognitive load induced by certain activities. Also, from our systematic review it is evident that one should distinguish between AR applications in which both, the virtual and real-world elements are relevant for physics learning (e.g., magnetic field lines overlaid over a real-world magnet), and AR applications in which the real-world environment does not directly contribute to physics learning (e.g., showing planetary motion above a real-world table). From a learning sciences perspective, the latter type of AR applications is very similar to traditional simulations, only with a slightly more prominent feeling about the simulated phenomenon as being part of the real-world (i.e., more prominent feeling of presence).

Generally, when it comes to research designs that had been implemented in AR-studies from our article pool, mixed design and (quasi)-experimental design were used most frequently, followed by case studies and interpretative research. In general, we can see that qualitative research has been implemented less frequently than quantitative research.

This finding is in line with the mere fact that quantitative methods are still prevalent in physics education research, which is rooted in traditional physics training which mainly relies on quantitative data [123]. However, it should be noted that it is generally recommended to start investigating a topic (e.g., the effects of AR) by using qualitative methods [124]. In fact, the use of qualitative methods could allow us to gain a deep insight into the learning processes that are promoted by the use of AR technology, and this insight could help us to form more reasonable hypotheses for conducting quantitative research. We have found that most studies combined several types of data collection methods to evaluate the effects of AR in teaching physics topics. It is evident that, similar to other areas of physics education research, a written assessment was the most common method of data collection. However, there were also a considerable number of studies in which the observation method has been used, which is particularly effective for assessing learning outcomes related to competence in using the scientific method [125]. There were also two studies in which electroencephalography (EEG) has been used, for measuring students interest and attention [40,119]. Finally, an interesting recent development is related to researchers using deep learning algorithms for recognition of real-world objects and their integration with corresponding virtual objects [57].

## 3.2. What are the specific advantages and disadvantages of using AR in teaching about physics topics?

Next, we will analyze possible advantages and disadvantages of AR-based physics learning activities. Our analysis is based on findings from the 67 empirical research articles that we included in our systematic review. A meta-analysis of the AR effects has not been performed because the article pool includes very heterogenous research contexts and designs.

### 3.2.1. Potential advantages

Figure 2 shows how frequently certain potential advantages of AR were mentioned in the identified 67 empirical research articles.

For a more detailed specification of the advantages discovered in each empirical research article, see Electronical Supplement A.

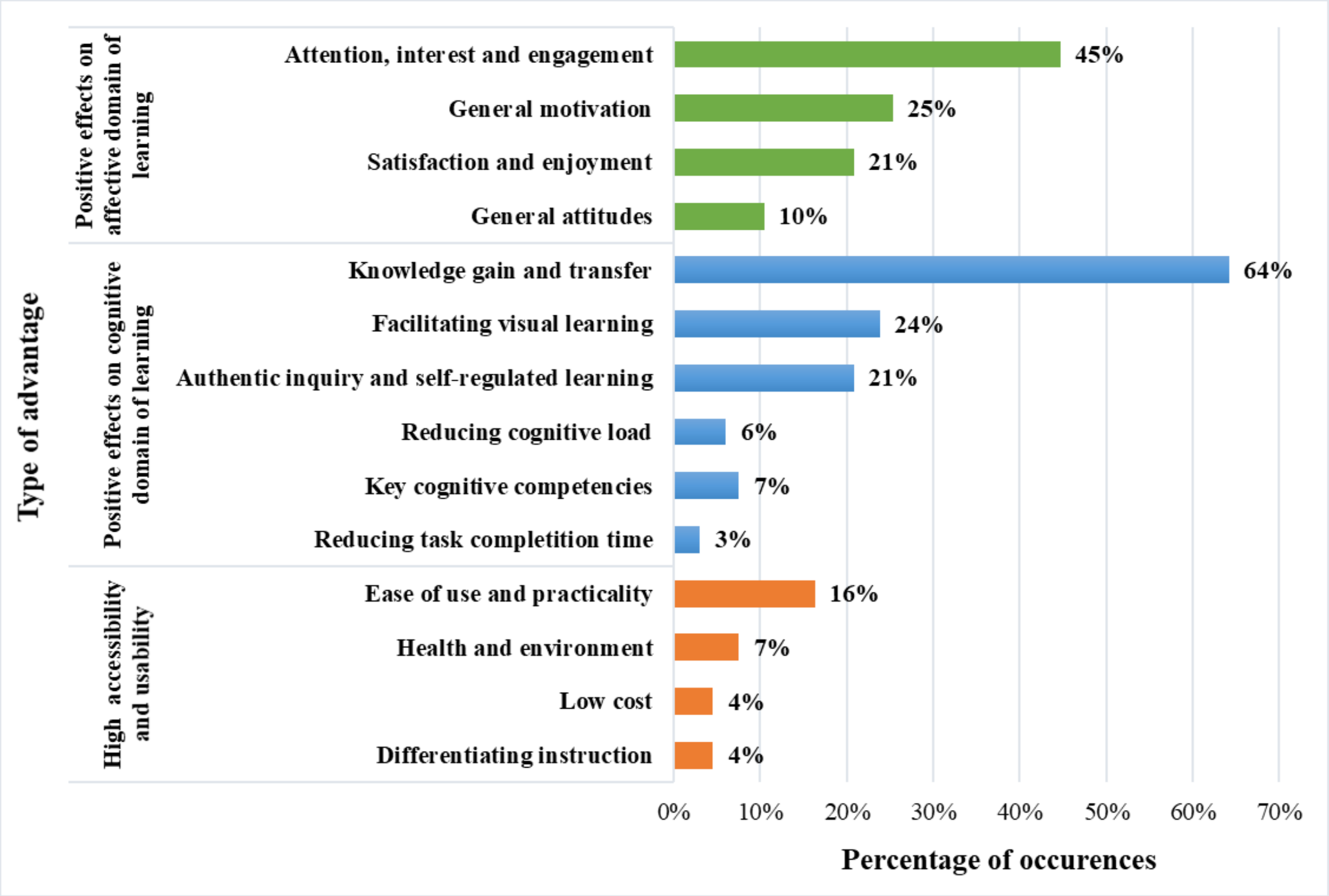

Figure. 2. Advantages of using AR in teaching physics; non-mutually exclusive analysis categories were used because in certain articles multiple advantages have been mentioned.

Next, we provide a discussion about the individual categories of identified AR advantages.

## 1. Positive effects on affective domain of learning

Many of the identified empirical research articles point out the positive effects of AR on students' affective domain of learning. This is an important result because affective aspects of learning influence not only students' cognitive achievement in physics, but also guide their academic path and influence their career choice [126]. In fact, learning cannot begin if there is no motivation for learning. Here, we can define motivation as "an internal state that arouses, directs, and maintains behavior" [127, p.660]. Content-specific motivation is often described through the construct of "interest" which is defined by Schiefele as "content-specific motivational characteristic composed of intrinsic feeling-related and value-related valences" [128, p.299]. Student's interest, active participation, attention and curiosity, may be described by the umbrella concept of student engagement [123]. A state of students' optimal experience characterized by full immersion and engagement in an activity, often resulting in a loss of time awareness and self-consciousness, is denoted as "flow". This mental state is marked by complete focus, enjoyment, and a perception of control over one's actions [129]. Whether or not students will reach high levels of engagement and a state of flow may be affected by their attitudes towards school physics and their sense of self-efficacy. Here attitudes may be defined as "feelings, beliefs and values held about an object" [130, p.1053], while self-efficacy is a "person's sense of being able to deal effectively with a particular task" [127, p.662].

Conclusions about students' affective domain of learning in the context of AR have been drawn through interviews, written surveys, observations, and even EEG measurements. It has been found that students who learned with AR were more intensively engaged in classroom discussions [36,39,77,85,117,120,122,131], were more focused [10,54,78,110,117], showed more enjoyment and interest [10,36,43,46,54,62,68,75,83,85,89,105,112,115,117,119,131] and more often overcame prejudices against the course [36]. Additionally, in the study by Arici *et al.* the teacher reported that students liked to learn within an AR environment because it reminded them of a game environment [36]. Some studies have reported students' prolonged motivation for learning based on their prior AR learning experience [122]. Therefore, it is not surprising that students and teachers in some studies expressed their positive attitudes towards AR, their willingness to continue using AR in the future and recommending it to others [36,48,85,106,119,121]. Alvarez-Marin *et al.* found that students were willing to use AR apps if they found them useful, not just easy to use [34]. The use of AR technology also proved to be promising when it comes to developing desirable student attitudes towards a physics course, in general [39,95]. This could be related to the construct of self-efficacy. Specifically, Chang and Hwang found large between-group differences in favor of the AR group when it comes to the construct of self-efficacy [45].

Particularly strong effects on affective domain of learning have been observed in studies related to learning about astronomy and electrical circuits [37,51,61,89,105]. Cai et al. [132] and Chen et al. [46] reported that students were able to enhance their flow experience when utilizing AR applications for learning about mechanics and astronomy. It should also be noted that students were more excited about applying smartphones on markers and observing visualizations, rather than watching the same animations on computers [49].

Next, it is useful to discuss why AR is perceived positively by many students. One possible factor that could account for some of the positive perceptions is that the use of AR may lead to a higher sense of control in students [64]. Another important factor could be related to the novelty effect that students associate with these technologies [42]. In fact, the novelty effect is known to trigger students' motivation and promote their initial interest [133-137].

## 2. Positive effects on cognitive domain of learning

Next, we will discuss some of the possible explanations for AR's positive effects on the cognitive domain of learning physics. In the identified literature, the positive effects of AR are most often explained by its potential to reduce task completion time, reduce cognitive load, facilitate visual learning, and make collaboration and inquiry more effective and realistic.

### a. Authentic inquiry and self-regulated learning

Social interaction, self-paced learning, and active interaction with learning material are critical to learning physics [138]. AR can provide a challenging visual context that facilitates communication/collaboration, which in turn can enhance physics learning. Learning can also be enhanced by the interactivity feature of AR, which facilitates the implementation of science inquiry processes and allows learners to explore physics phenomena safely and at their own pace.

Next, we provide support for the above-mentioned claims. Since the pedagogical potential of external representations is highly content-dependent [139], we attempt here to present our findings separately for different areas of physics.

*Astronomy*

Çakıroğlu *et al.* [43] reported that AR helped students to create concept maps more accurately and faster by providing instant feedback to users when creating concept maps. In other words, AR-generated feedback may help the learners to more effectively learn in the absence of teacher-feedback.

*Electricity and magnetism*

In Abdusselam & Karal's study, students reported that the AR-assisted laboratory was a more realistic environment for activities of shape visualization and learning about magnetism concepts [31]. It should also be noted that many students related the positive effects of AR to its interactivity [92]. In the study by Chang & Hwang students pointed out the positive experience of being able to practice at their own pace and having personalized operation [45]. Finally, in the study by Ibáñez et al. for the AR group higher levels of clear feedback and autotelic (internally driven, intrinsically motivated) experience were reported than for the control group [63]. Students' interviews and video footage from S. A. Yoon *et al.*'s [93] study about electrical conductivity and circuits showed that students identified collaboration in AR group as the most helpful scaffold that helped them fully understand the flow of electricity and the different ways a complete circuit can be constructed.

In the study by Mukhtarkyzy *et al.* [83] an interactive scenario generated by the AR application enabled students to manipulate circuits and their components in real-time, similar to practical work conducted in a traditional laboratory.

*Mechanics and mechanical waves*

In the study about elastic collisions by Wang *et al.* [60], it was shown that in using AR, students' learning behavior patterns were similar to the patterns which characterize ideal collaborative inquiry processes. The feature of providing a more realistic environment for inquiry has been also mentioned in the context of mechanics [59]. Some researchers pointed out that the AR system supports knowledge construction by real-time data collection and representation [71]. Cai *et al.* [132] reported that the utilization of a Brain-Computer Interface (BCI)-based Augmented Reality (AR) inquiry tool, incorporating EEG technology, had a significant positive impact on students' scientific performance and their understanding of the lever principle. The use of BCI-based AR, coupled with continuous positive and negative attention feedback, played a crucial role in ensuring that students consistently maintained their attention levels above a specific threshold.

### b. Facilitating visual learning

Visual learning is a type of learning that focuses on visual elements incorporated in the lessons [140]. This type of learning is very important in physics education, especially for learning topics that require intensive spatial reasoning, such as: astronomy, electromagnetism and optics. Augmented reality technology has the potential to positively influence visual learning through carefully designed auxiliary visualizations that can enrich authentic representations of physical objects and phenomena.

Next, we provide support for the above-mentioned claims.

*Astronomy*

In the study about lunar phases by Tian *et al.* [109], it has been concluded that the AR application was very useful in facilitating visual learning and developing understanding of the relationships between celestial objects. Similarly, in the study by Phon *et al.* [88] about learning astronomical topics, significant learning gains were found for the AR group and the gains were significantly larger for low-spatial ability than for high-spatial-ability students. Arici *et al.* found that students think that AR technology helped them to reduce abstractedness of knowledge and to observe structures (celestial bodies) that are otherwise unobservable in the classroom environment [36]. Çakıroğlu *et al.* [43] reported that in creating concept maps about lunar and solar eclipse, AR-generated visualizations triggered schemas which helped the students to connect different elements of knowledge. In the study by Önal and Önal students stated that visualization helped them to more easily learn astronomical topics [85]. Please note that in this study only students' "feeling of learning" has been measured, which does not necessarily match the actual learning [141]. In Chen et al.'s study, students mentioned that the incorporation of a 3D animation to depict the simulated motion of real planets significantly facilitated discussions with their peers about topics they found challenging and allowed them to revisit and review the learning content [46].

*Atomic and molecular physics*

In the study on learning about the photoelectric effect by Cai *et al.* [42] students reported that the AR application helped them to better visualize electron movement. Similarly, Majid & Majid [77] claimed that visualizing the 3D model of each atom allowed their students to gain more conceptual understanding about atomic structure. Finally, Arici *et al.* asserted that the AR application helped students to internally visualize the unobserevable structure of the atoms [36], and in the study by Keifert *et al.* [68] the AR-based visualization was supposed to establish a connection between students' imaginative embodiment of particles in motion and conceptual ideas they had about properties of states of matter.

*Electricity and magnetism*

In their study on AR-based learning about electromagnetism, Ibáñez *et al.* [63] found that AR-group students had significantly higher post-treatment scores for visualization than control group students who learned with a web-based application. Furthermore, information from interviews in the study by Liu *et al*. showed that AR helped students to visualize abstract knowledge by superimposing virtual objects (magnetic field lines, geomagnetic field) onto a real world scene [75]. Seo and So showed that gesture based AR (e.g., showing specific visualizations in response to corresponding gestures) may help in conceptual learning about electric circuits [98]. Similarly, in the study about electrostatic phenomena by Wang, the teacher who implemented the educational intervention stated that AR materials helped the students to associate visual learning information with concrete physical objects, which made it easier for them to complete the assigned tasks [120]. Sonntag and Bodensiek showed that the AR application helped in developing conceptual understanding by guiding the students' visual attention [99].

Finally, in their research about electromagnetic waves, Faridi *et al*. concluded that students from the AR group were able to better visualize abstract concepts such as magnetic field, current flow, and effects of an increase in electric potential [58].

### c. Key cognitive competencies

In this context, key competencies can be defined as competencies necessary for success in daily life and for a well-functioning society [142]. When it comes to key cognitive competencies, it has been shown that AR promotes critical thinking skills [45,58], and abstract thinking [36,103]. Clearly, more studies are needed to obtain stable conclusions about the effects of AR-based learning on the development of key cognitive competencies.

### d. Optimizing cognitive load

Only three studies from this review explicitly included measurements of cognitive load. Findings from these studies suggest that augmented reality technology could help to optimize cognitive load and improve learning of the selected topics in Astronomy, Magnetism, and Thermodynamics and heat. There have also been some studies from other areas, such as electrical circuits, for which there is some indirect support for the argument that AR technology has helped to reduce irrelevant cognitive load.

Within the context of astronomy, Liou *et al*. [74] suggested that real objects presented in the AR system helped students to optimize cognitive load because students could take real objects in the AR system as reference objects of the moon movement, which potentially increased relevant cognitive load. Information about the date, time, direction, elevation, shape of the moon, and real objects were displayed on the screen. Students who used the AR system with integrated multimedia elements had positive emotions, which may also have contributed to the optimization of cognitive load through the mechanism of motivation to invest larger efforts.

When it comes to electrical circuits, in the study by Ibanez *et al*. [64] some students from the AR group indicated that using AR application helped them learn about circuits with less effort.

For the content area of magnetism, Liu *et al.* found that the AR-application from their study helped the students to reduce cognitive load, by providing a virtual-real integrated environment which facilitated construction of knowledge [75].

Finally, in their study about heat conduction in metals, Thees *et al.* [108] found that students in the AR group exhibited a significantly lower extraneous (irrelevant) cognitive load than their peers in the traditional control group.

### e. Reducing task completion time

Besides the goal to develop certain experimental skills, another important goal of physics labs is often related to developing conceptual understanding of physics. Reducing task completion time may largely facilitate learning, particularly in the laboratory context, where accelerating completion of technical aspects of an experiment helps the students to focus on conceptual aspects of a particular activity. Support for the claim that AR technology potentially helps in reducing task completion time, has been obtained for the contexts of electrical circuits and mechanics/mechanical waves. Specifically, through interviews, it was shown that the AR application helped to reduce the instructor's workload in the laboratory because the application proved to be effective in providing students with relevant information about the experiments, i.e., providing them with implicit guidance [143]. Additionally, Çakıroğlu *et al.* [43] noted that AR technology facilitated faster task completion by providing the students with instant feedback on their work.

In the study by Costa *et al.* teachers reported that the AR application proved to be useful for assessing students' performance during the game activity, i.e., that it has potential to make evaluation of students' achievement a more time-efficient process [48].

## 3. High accessibility and usability

An educational technology is more probable to have a major impact on an educational system if it is easily accessible and practical.

### a. Ease of use and practicality

The inherent complexity of learning aids and materials can directly affect learning by increasing extraneous (irrelevant) cognitive load. It is important that students feel comfortable with educational technologies so that they can fully enjoy learning. Previous research has shown that many students think that AR is practical and easy to use.

For example, in the study by Reyes-Aviles & Aviles-Cruz, students even stated that they valued AR systems more than traditional measurement instruments [90]. These findings were supported by the study of Altmeyer *et al.* [33], who found that students especially appreciate the fact that errors can be resolved quickly and easily in AR physics labs. Students also frequently mention that AR is easy to use. In particular, Tomara & Gouscos [110] asserted that students had no difficulty using mobile devices during the AR activity. They distinguished between real objects and augmented data with relative ease and observed digital data overlaying augmented targets with little effort [110]. That AR is easy to use was also mentioned by students in the studies by Tian *et al.* [109], Tarng *et al.* [106], Xiao *et al.* [119],

Liu *et al.* [75], Önal and Önal [85]. Oh *et al.* [84] point out that wearable AR glasses are particularly practical and effectively support embodiment because users do not need to hold a mobile display for augmentation. In Chen *et al.*'s [46] research on planetary motion, students claimed that the Augmented Reality (AR) system was very easy to operate. Moreover, in the study conducted by Ropawandi *et al.* [131], students who were interviewed also highlighted the suitability, convenience, accuracy, and usefulness of the AR application for learning during the pandemic. They asserted that the AR application offered a convenient solution for their online learning, particularly in addressing misconceptions and facilitating a more profound comprehension of physics concepts.

Finally, an important conclusion by Majid and Majid [77] is that because of students' willingness to use the AR application in their free time, we can say that AR leads to flexibility of learning, i.e., it becomes possible to learn anytime and anywhere.

### b. Low cost

Sometimes the cost of equipment is a limiting factor in setting up physics education laboratories. Although generally, traditional hands-on laboratories cannot and should not be completely replaced by AR-based experiments, in some occasions AR technology can provide an affordable, relatively effective alternative to expensive traditional labs. As always, traditional and AR-based experiments may function complementary to facilitate development of a wide range of competencies.

For example, the low-cost AR tool presented by Cai *et al.* [42] solves the difficulties associated with unfavorable light source control and the low accuracy of ammeters in ordinary classrooms, which helps students visualize and learn abstract concepts related to the photoelectric effect. In another study, Cai *et al.* [41] pointed out that AR tools can substitute expensive and complex devices. For example, the plasma discharges experiment can be performed at a fraction of the cost of a real experiment when using the augmented reality application (90 vs. 20 000 AUD) [121].

### c. Health and environment

In the modern educational process, great attention is paid to creating a safe learning environment. This particularly holds for physics learning which may include activities that potentially represent a risk for teachers' and students' health. Consequently, discussing approaches to reduce health risks are very important. One of these approaches is the use of AR in our classrooms.

For example, in the study about Sun path by Tarng *et al.* [106], students and teacher asserted that using the AR system could prevent the risk of heatstroke caused by direct sunlight while solving tasks outdoors for an extended period of time. In addition, the AR-approach to studying Sun path may also overcome difficulties associated with bad weather or topographical constraints that often make observation difficult.

Furthermore, teachers from the study by Arici *et al.* [36] and students from the study by Liu *et al.* [75] and Önal and Önal [85], pointed out that AR technology prevents dangerous situations that could arise in the wide range of experiments conducted in real classroom environments. Finally, Yau *et al.* concluded that health risks from high voltage current flowing through the body can be avoided if AR technology is used to create an experiment that involves discharges of plasma [121]. Similar conclusions can be drawn for experiments involving electrical circuits that operate at high voltages.

Finally, it is important to note that replacing real-world experiments with AR-based experiments may also have some drawbacks, not only related to development of certain experimental skills (e.g., data collection and processing), but also when it comes to the epistemological perspective (e.g., physics knowledge in its essence comes from us observing the real world).

### d. Differentiating instruction

Previous research has shown that using the same teaching methods and materials can lead to different learning outcomes for students of different ages and genders [144-146]. Therefore, it is important to report on how AR affects learning in different student populations.

In their study about electrostatic forces between point charges, Echeverría *et al.* [55] found that AR benefited significantly more boys than girls. Similarly, in the study by Suprapto *et al.* [104] boys outperformed girls in learning about planet motion.

Although these initial studies indicate that AR technologies may be more beneficial for boys than for girls, further studies in multiple educational contexts are needed, in order to provide reliable suggestions related to using AR for differentiating the physics teaching practice with respect to gender.

When it comes to a possible interaction with the age variable, in the study by Urbano *et al.* [112] it has been shown that older students respond in a similar way to AR as younger ones.

### 3.2.2. Potential disadvantages

Every educational technology has potential disadvantages that, under certain circumstances, may negatively influence the process of learning. Figure 3 shows how frequently certain AR-based learning disadvantages were reported in our 67 empirical research articles.

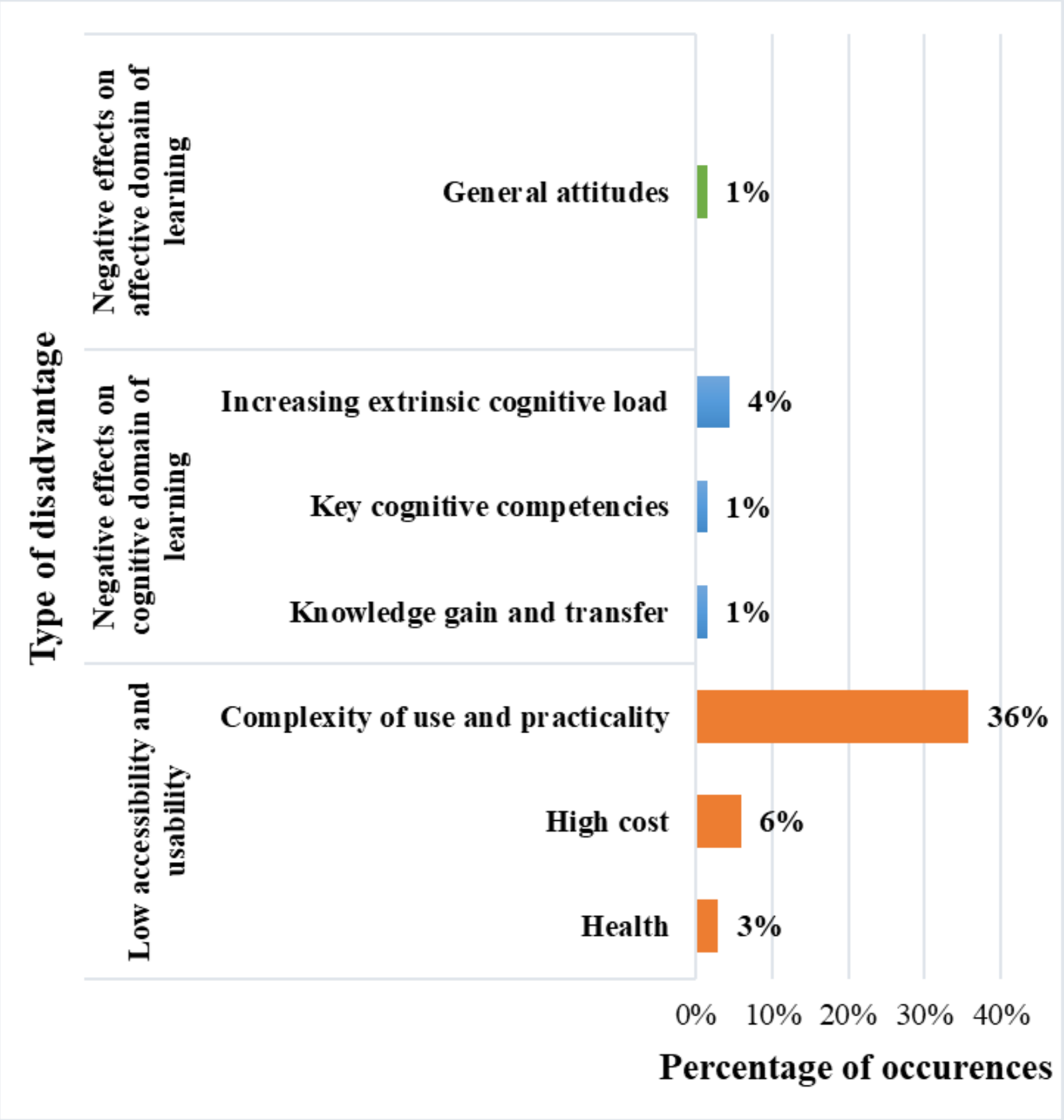

Figure 3. Disadvantages of using AR in teaching physics; non-mutually exclusive analysis categories were used because in some articles more than one type of disadvantage has been reported.

For a more detailed specification of the disadvantages discovered in each empirical research article, see Electronical Supplement B.

Next, more details about the various types of potential disadvantages of AR-based learning of physics will be provided.

## 1. Negative effects on affective domain of learning

In only one study negative effects of AR technologies on the affective domain of learning have been reported. Concretely, in the research by Coşkun and Koç students who learned with AR technology showed significantly more expressed negative attitudes towards learning about astronomy topics compared to students from the traditional group. Furthermore, for the AR-based group anxiety levels were significantly increased in comparison with the pretreatment values [47]. Coşkun and Koç tried to explain these findings by referring to students' unease with being object of research, unfamiliarity with the new educational situation and fear of academic failure.

## 2. Low accessibility and usability

### a. Complexity of use and practicality

In some of the AR studies, students gave low ratings to the technical usability of AR [119]. The reported issues were mainly related to imperfections of software or hardware technology. Students reported software problems that are mostly associated with deficient AR tool designs: interaction instability [42], inadequate user interface characterized by small font size or difficulty in entering numbers [109,110], low image and sound quality [105], application navigation difficulties [75], issues with object tracking [115], detection failure [69,117] and slow software leading to delays [147].

On the other hand, teachers had problems with time management and more demanding planning and implementation of the teaching process, e.g., setting up Wi-Fi and AR equipment [120]. Echeverría *et al.* pointed out that computer simulations were easier to manage than AR and that in the AR environment it was more complicated for the teacher to follow the learning process and help the students [55]. Costa *et al.* reported that teachers needed technical guidance from the facilitators to register and use the tools provided by the AR system [48]. Due to high technical requirements, the researchers in Cai *et al.*'s [9] study hired assistants who were responsible for technical questions and explaining the operation procedures.

When it comes to reported hardware problems, they were mainly related to the complicated setup and inadequate hardware devices. For example, some researchers reported difficulties due to camera freezing and increased energy consumption [36,69]. Other researchers reported that their application was unable to register all image markers simultaneously, due to the slow processing speed of the used smartphones or tablet devices [121]. It has also been reported that marker-based AR can be affected by various features of light exposure which may lead to recognition error [8,36,41,45,117]. Furthermore, in the study by Önal and Önal students complained about small mobile device screens while in the study by Çakıroğlu *et al.* students reported certain difficulties in learning with the AR, attributing these difficulties to the way they had to hold their mobile phones [43,85].

In studies in which AR glasses were used, sometimes the limited field of view, together with bulky and heavy glasses, has been mentioned as one of the main difficulties [84,101].

### b. Health

Potential health risks were detected by Fidan and Tuncel [59] who reported that students from the AR group experienced eye problems, neck stiffness, and pain in their arms and hands from holding and manipulating the tablet for a relatively long period of time. One student from the study by Düzyol *et al.* [54] stated that for her/him it was difficult to hold the tablet in the specified way while being engaged in learning activities.

### c. High cost

Echeverría *et al.* [55] reported that the AR setup was more expensive than the computer simulation setup. Specifically, the AR setup required one appropriate tablet device per student, while the computer simulation setup required one laptop per three children. In the study by Arici *et al.* [36] teachers reported concerns about the high cost of an AR system that included specific tablets and paid mobile applications while Önal and Önal reported that some participants raised concerns about the high cost of mobile devices and price of the used AR system [85]. In the study by Talan *et al.* [105], students complained about the high cost of the AR application and cards (e.g., Space 4D +) employed in their research.

It is important to note that, similar to computer-simulations, there are free to download AR apps, as well as paid ones. Certainly, if specialized hardware equipment is needed (e.g., head-mounted displays) the costs of AR-based learning additionally rise.

## 3. Negative effects on cognitive domain of learning

There are only a small number of studies in which non-significant or negative effects of AR on learning about physics phenomena were reported. However, we must be cautious in interpreting this finding, because it is well known that journals tend to publish positive, statistically significant research results. In other words, there were probably more studies that found non-significant differences between control and AR groups, but these articles were never published in journals indexed by Scopus or Eric.

Also, early implementers of technologies such as AR tend to be enthusiasts which increases the probability of obtaining positive, statistically significant findings. Consequently, these findings may not necessarily transfer to other contexts with less enthusiastic/skilled teachers.

### a. Key cognitive competencies

In the article by Wang [120], it was pointed out that AR-content might limit students' creativity because some students strictly follow the AR generated instructions without thinking deeper about the mere physics task. However, this potential shortcoming can probably be avoided by careful design of the AR learning activity and cannot be perceived as an inherent shortcoming of the AR technology.

### b. Increasing extraneous cognitive load

Details that are part of the learning environment but irrelevant to learning the target knowledge negatively influence learning. This is known as irrelevant (extraneous) cognitive load. Baran *et al.* [38] observed that during AR-based group learning about electrical circuits frequently conflicts occurred because almost every student in the group wanted to have the AR-devices in her/his own hands. Consequently, some students stated that they could not observe the target phenomena. In the same study, students showed a tendency to talk about how to use the AR-devices, instead of focusing on the work tasks.

Ibáñez [63] reported difficulties related to the simultaneous handling of the tablet and the physical objects that are representing the components of the electrical circuit, which led to cognitive overload for some students. There is particularly high risk of cognitive overload when there is a lack of representational guidance to support learners' cognitive behaviors.

### c. Knowledge gain and transfer

Cai *et al*. [9] reported that immediately after the learning intervention, students from the traditional groups outperformed their colleagues from the AR group in understanding magnetic field laws. However, this difference was negligible on the repeated test one week after the initial experiment.

# 4. Conclusions

In this paper, we presented the findings from a systematic review of literature that aimed to identify opportunities and challenges in AR-based learning of physics. Our review included articles published between January 1st, 2012 and January 1st, 2023 that were indexed in either the Scopus or Eric databases and discussed the use of AR in learning about physics.

When it comes to the ways in which AR is utilized for facilitating learning of physics, we could generally conclude that AR-applications may be used to:

- enrich the experience of conducting real-world experiments by overlaying visual representations of abstract mechanisms related to the observed physics phenomena (e.g., by blending in content-specific visualizations such as light rays or by visualizing underlying microscopic processes),
- effectively organize learning activities by providing relevant textual/audio/video on top of real-world scenery (e.g., instructions on how to conduct experiments or use certain equipment)
- facilitate kinesthetic learning which often includes haptic experiences and exploring cause and effect relationships (e.g., by moving something in the real-world, certain changes in the virtual world occur)
- enrich the real-world scenery with data obtained from sensors (e.g., color-represented temperature values are overlaid on top of real-world scenery)
- improve textual learning materials, by overlaying corresponding 3D-visualizations which potentially helps to better understand the text (e.g., providing a 3D-visualization of moon phases to complement the textual description).

We can also conclude that, provided their design and use are properly aligned with corresponding learning goals, AR technologies may have a positive effect on cognitive and affective domain of learning physics. Thereby, the positive effects of AR are most often explained by its potential to:

- reduce task completion time (e.g., by reducing the time for completion of some technical aspects of an activity, more time remains for discussion),
- optimize cognitive load (e.g., all instructions on how to operate certain pieces of experimental equipment are effectively spatially organized on top of real-world scenery, so the learner has not to split their attention on looking at different places for finding instructions),

- facilitate visual learning (e.g., visual representations are integrated with abstract representations into functional mental models of physical phenomena),
- make collaboration and inquiry more effective and authentic (e.g., providing a visual context situated within real-world scenery facilitates communication/collaboration about physical phenomena, which in turn enhances physics learning)
- provide the students a higher sense of control over their own learning (e.g., AR application generates specific content/feedback when required by the user).

Other important advantages of AR are related to the fact that in certain situations AR technologies can lower health risks (e.g., some AR experiments may be safer to conduct than corresponding real-equipment experiments) and make experimenting experiences possible (e.g., when the real equipment is very expensive).

When it comes to potential challenges of AR based learning, some authors have reported that AR based learning may be associated with increased extraneous cognitive load (e.g., struggling to understand technical aspects of the application, simultaneous handling of AR hardware and physical objects) and increased time spent-off task (e.g., talking about the mere technology instead of focusing on learning physics). Many of the reported challenges were of technical nature. One of the major software problems was image detection failure and slow software, while the hardware problems were mainly related to the complicated setup and inadequate hardware devices (e.g., camera freezing and power consumption).

## 5. Suggestions for teaching and research practice

In our paper, we have discussed the ways in which AR technologies may facilitate learning of physics, their potential advantages and disadvantages. We also briefly reflected on the research designs in earlier AR research. Next, we propose a few suggestions that may help to improve the existing teaching and research practices.

When it comes to the teaching practice:

- Teachers should take as much time as necessary to carefully select software and hardware devices that meet their specific needs, technical capabilities, and learning objectives.
- Teachers and AR developers are advised to pay more attention to the quality of AR application interface design (e.g., image and font size, colors, representational guidance, ease of use).
- When planning the implementation of AR lectures, physics teachers are advised to carefully consider how to avoid technical difficulties reported in earlier research, such as camera freezing and sensitivity, marker detection, memory and power consumption, GPS errors, etc.

- It is recommended to secure enough AR devices in the classroom, so that all students can actively participate in the learning process.
- Teachers are recommended to incorporate breaks or energy shifts into lectures to avoid students' fatigue when using AR while learning. Furthermore, they are advised to avoid using of poor quality AR glasses (limited field of view, bulky and heavy glasses).

Next, we provide suggestions for physics education researchers. In future AR research, it would be interesting to:

- explicitly compare different types of AR technologies, implemented for teaching about a single topic (e.g., AR glasses vs. mobile devices when teaching about lunar phases or marker-based vs. markerless approach when teaching about the gravitational law).
- investigate whether the effectiveness of AR depends on certain students' characteristics, such as socioeconomic background, visual reasoning ability, and technical knowledge.
- conduct long-term studies lasting for several months to investigate the impact of the novelty effect, for different population groups.
- thoroughly investigate ways for optimizing the cognitive load in AR-based learning which should also include measurements of cognitive load.
- conduct more in-depth, qualitative studies about the students' experience with AR in learning physics.

Finally, following the example of the Phet repository, it would be useful to create a repository of AR applications along with guidelines for their implementation in physics teaching practice.

# 6. Limitations of the study

In this systematic review, we have only considered articles from the Scopus and Eric databases. We did not perform searches in other important databases, such as Web of Science or GoogleScholar. GoogleScholar has been avoided because it is widely known to also index articles from low quality journals and such articles may offer misleading conclusions about AR in physics learning. On the other hand, not searching the Web of Science database probably did not significantly affect the quality of our review, because the majority of articles that are indexed in Web of Science are also indexed in Scopus or Eric. In fact, an analysis performed on October 3rd 2023, showed that 268 out of the 269 journals indexed in Web of Science - SSCI (Education & Educational Research) were also indexed by Scopus or Eric.

Some relevant articles may have been also missed because we performed our database searches by using only the terms "augmented reality" and "mixed reality". Future review studies should additionally use "assisted reality", "extended reality" and other related terms, in order to identify a larger number of potentially relevant AR articles. Nevertheless, it is important to point out that this systematic review was based on a larger number of articles compared to many recent AR-reviews that have attempted to cover a much broader field (e.g., the field of STEM education).

In addition, we included in our review only peer-reviewed journal articles written in English. In future studies, researchers could consider to also include other types of high-quality publications (e.g., conference papers, dissertations and theses, technical reports) published in multiple languages. This could help reduce the issue of publication bias, which is reflected in the fact that journals are more inclined to publish articles that report statistically significant between-group differences.

Another limitation of this research is that we did not conduct rigorous inter-coder agreement studies. We believe that this did not significantly compromise the objectivity of the article selection process, as this process mainly required low-inference decisions. A similar observation holds for classifying articles with respect to types of AR-related advantages and disadvantages.

**Acknowledgments**

We are very grateful to the anonymous reviewers whose insightful comments and concrete suggestions helped us to improve the quality of our manuscript considerably.